\documentclass[%
 reprint,
superscriptaddress,
 amsmath,amssymb,
 aps,
prb,
onecolumn
]{revtex4-2}
\usepackage{subcaption}
\usepackage{graphicx}
\usepackage{dcolumn}
\usepackage{bm}
\usepackage{caption}
\usepackage{blindtext}
\usepackage{physics}
\usepackage{xcolor}
\usepackage{tikz}
\begin{document}

\preprint{APS/123-QED}

\title{Autoregressive Typical Thermal States}

\author{Tarun Advaith Kumar}
\affiliation{Department of Physics and Astronomy, University of Waterloo, Ontario, N2L 3G1, Canada}
\affiliation{Perimeter Institute for Theoretical Physics, Waterloo, ON N2L 2Y5, Canada}
\author{Leon Balents}
\affiliation{Kavli Institute for Theoretical Physics, University of California, Santa Barbara, California 93106-4030, USA}
\affiliation{Canadian Institute for Advanced Research, Toronto, Ontario, Canada}
\author{Timothy H. Hsieh}
\affiliation{Perimeter Institute for Theoretical Physics, Waterloo, ON N2L 2Y5, Canada}
\author{Roger G. Melko}
\affiliation{Department of Physics and Astronomy, University of Waterloo, Ontario, N2L 3G1, Canada}
\affiliation{Perimeter Institute for Theoretical Physics, Waterloo, ON N2L 2Y5, Canada}

\date{\today}

\begin{abstract}

A variety of generative neural networks recently adopted from machine learning have provided promising strategies for studying quantum matter.
In particular, the success of autoregressive models in natural language processing has motivated their use as variational ans\"atze, with the hope that their demonstrated ability to scale will transfer to simulations of quantum many-body systems.
In this paper, we introduce an autoregressive framework to calculate finite-temperature properties of a quantum system based on the imaginary-time evolution of an ensemble of pure states.
We find that established approaches based on minimally entangled typical thermal states (METTS) have numerical instabilities when an autoregressive recurrent neural network is used as the variational ans\"atz.
We show that these instabilities can be mitigated by evolving the initial ensemble states with a unitary operation, along with applying a threshold to curb runaway evolution of ensemble members.
By comparing our algorithm to exact results for the spin 1/2 quantum XY chain, we demonstrate that autoregressive typical thermal states are capable of accurately calculating thermal observables.
\end{abstract} 

\maketitle


\section{Introduction}

Autoregressive neural networks have become one of the most prominent architectures used by the generative artificial intelligence community. 
The basic functionality of these networks is likelihood estimation, which includes modeling a target distribution by learning a set of parameters, and also performing inference to generate new samples.
Autoregressive models rely on decomposing a joint distribution over multiple variables into a sequence of conditional distributions,
\begin{equation} \label{Pjoint}
    p({\bm x}) = \prod_{i=1}^N p(x_i | {\bm x}_{<i}).
\end{equation}
The variables $x_i$ could be $N$ words of a corpus, or the individual projective measurement outcomes from $N$ qubits say. 
Whatever the application, through this decomposition into a sequence, autoregressive models encode a normalized parameterization of the joint distribution $p({\bm x})$ across the $N$ variables.  Autoregressive models can enable a number of efficiencies during learning and inference, such as direct sampling without the need for Markov chains.

The first autoregressive models included the
Fully Visible Sigmoid Belief Network (FVSBN) without any latent space, and the 
Neural Autoregressive Density Estimation (NADE) 
which provided a way to vary the expressiveness through a hidden layer.
Regardless of the architecture, autoregressive models necessarily encode all of the dependencies through a sequence (Eq.~\ref{Pjoint}).
Early architectures, such as recurrent neural networks (RNNs) which maintain a hidden state, struggled to encode long-range dependencies between variables separated by large distances in the sequence \cite{bengio1}.  This hurdle was overcome with the long short-term memory (LSTM) in RNNs \cite{lstm}, and more recently by the self-attention mechanism in autoregressive transformer architechtures \cite{vaswani}. These developments have led to autoregressive models being widely adopted in natural language applications.
However, even with these advanced architectures, the training procedure can suffer from a high variance in the gradients and sensitivity to parameter initialization. Substantial effort has been directed towards developing heuristic improvements in training and initialization of autoregressive networks, leading to the boom in today's language models \cite{vanExpGradRnn, transformerInit}. 

More generally, the strong performance of neural networks, coupled with the accessibility provided by modern autograd frameworks, has driven their adoption for a wide variety of computational tasks in the physical sciences \cite{RevModPhys1}. Within physics, neural network architechtures have been particularly popular for quantum matter simulations \cite{jcMLforQM}. Framing quantum simulation as an optimization problem through the variational principle allows for neural network representations of quantum states to serve as an ansatz for variational Monte Carlo (VMC) \cite{Carleo1, J1J2, moss2025leveragingrecurrenceneuralnetwork}. 
Thus, neural network strategies provide a complement to more traditional methods such as quantum Monte Carlo (QMC) or tensor network simulations \cite{QMC1, QMC2, TN1}.
The hope is that neural networks might expand the class of quantum systems where numerical strategies can be applied, due to their high degrees of
expressivity and flexibility. 

Considerable effort has been devoted to developing representations of quantum states using specialized neural network topologies. Energy-based, autoregressive, and latent variable architectures have been effectively applied in ground-state searches, with notable success achieved by encoding symmetries directly into the ans\"atz 
\cite{SymmNQS1, SymmNQS2}. Extensions of the VMC paradigm have even been developed for simulating dynamics of quantum states. Neural network-based approaches have also been applied more recently to open quantum systems. Methods such as parameterization of the matrix elements of the density operator $\rho$ \cite{Saito, GHDO}, representing a purified state \cite{purify}, or the use of POVM ansätze have been explored \cite{POVM}. In the context of finite-temperature simulations, these need to be combined with imaginary-time evolution from infinite temperature, Lindbladian long-time simulation, or modified free-energy minimization to extract thermal observables \cite{NQSReview, Cirac}.  

One approach for representing the Gibbs state of a quantum Hamiltonian is the minimally entangled typical thermal state (METTS) algorithm first developed by White \cite{white}. This technique decomposes the density operator into an ensemble of pure states and evolves from an infinite-temperature state \cite{ Stoudenmire}. Hendry, Chen, and Feiguin demonstrated the efficacy of a neural-network-based METTS using an energy-based restricted Boltzmann machine (RBM) ansatz \cite{hendry}. A natural extension is to implement a similar approach with autoregressive architectures, which could offer advantages in regimes where sampling energy-based models becomes challenging. However, as we discuss in detail below, a key drawback is the numerical instability that can arise during the time evolution of a neural network state. These issues are particularly pronounced for autoregressive models, limiting their direct application in recreating METTS \cite{donatella, bengio1, vanExpGradRnn, fidelity, geoNQS}.

In this paper, we introduce modifications to the METTS algorithm aimed at mitigating instabilities arising from the time evolution of autoregressive states. To assess performance, we apply our approach to a solvable 1D quantum XY model, which enables direct comparison with exact results. We find that with methodical tweaks to improve the initialization and training dynamics of RNNs, autoregressive typical thermal states can faithfully estimate the thermal behavour of this XY model.

\begin{figure}[t]
    \centering
    \includegraphics[width=0.65\linewidth]{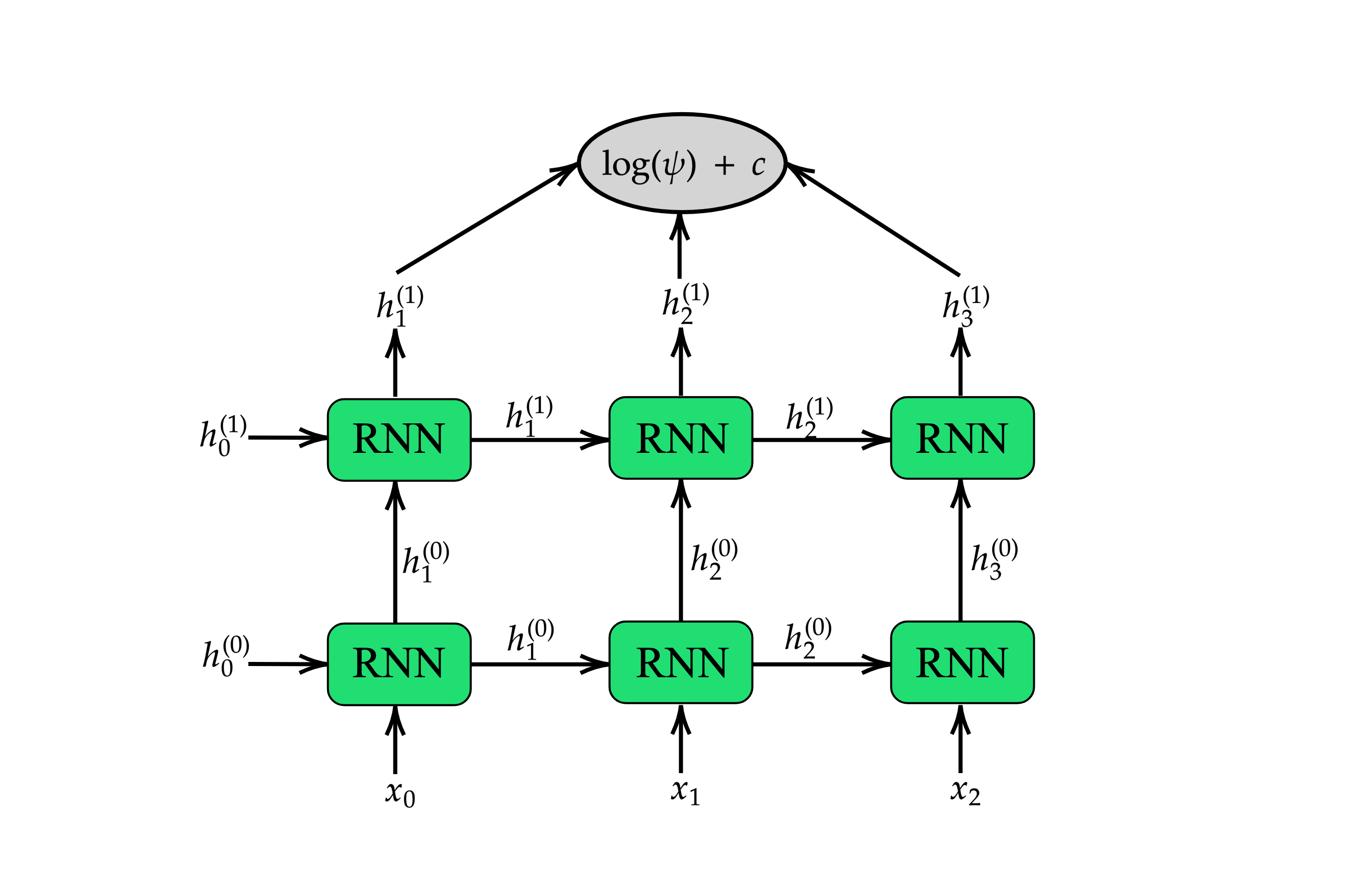}
    \caption{Our ansatz is a two-layer RNN with an LSTM cell. The output of the second layer LSTM cell is interpreted as the conditional log amplitude up to a normalization constant, and can be summed to form the log amplitude for a given spin configuration $\bm{x}$}
    \label{fig:ansatz}
\end{figure}
\section{METTS Theory}\label{sec:background}

A neural network can be used to parameterize a quantum state by considering it as a function $\psi_{\bm \theta}: X \rightarrow \mathbb{C}$. $X$ is the space described by the classical degrees of freedom and is referred to as the computational basis. Interpreting the function outputs as the wavefunction amplitudes in the computational basis gives us the state,
\begin{align}
    \ket{\psi_{\bm \theta}} \propto \sum_{\bm{x} \in X} \psi_{\bm \theta}(\bm{x}) \ket{\bm{x}} .
\end{align}
In analogy to Eq.~\eqref{Pjoint}, an autoregressive ans\"atz parameterizes the amplitude as a product of conditional amplitudes,
\begin{align}
    \psi_{\bm \theta}(\bm{x}) = \prod_{i=1}^N \psi_{\bm \theta}(x_i | {\bm x}_{<i}). \label{eq:psiJoint}
\end{align}
Time evolution of a state constrained to a variational manifold has been extensively studied -- the traditional approach is known as time-dependent variational principle (TDVP). It involves updating a variational state by minimizing the distance between the given state and an evolved version of itself \cite{TDVP1, TDVP2, TDVP3}. The main difference from steepest descent is the calculation of the distance using the Fubini-Study metric instead of the Cartesian one. In the case of real, positive wavefunctions, the Fubini-Study metric reduces to the Fisher-Information metric, and the following procedure is identical to natural gradient descent \cite{Amari, Martens}. TDVP requires the calculation of the quantum geometric tensor $S$ , 
\begin{equation}
    S_{\mu \nu} = \expval{(O_\mu - \expval{O_\mu})(O_\nu - \expval{O_\nu})} = \expval{\bar{O}_\mu \bar{O}_\nu} . \label{eq:S}
\end{equation}
Here the symbol $\expval{}$ represents the average over the spin configurations $\{\bm{x}\}$, and the notation $\bar{A}$ denotes a centered variable $\bar{A} \equiv A - \expval{A}$. For notational simplicity, let's assume a real, positive wavefunction and a variational ans\"atz with real parameters. $O_\alpha$ represents the logarithmic derivative of the ans\"atz with respect to a parameter $\theta_\alpha$,
\begin{equation}
    O_\alpha(\bm{x}) = \frac{\partial \ln \psi_{\bm \theta}({\bm x})}{\partial \theta_\alpha}; \,\, \mathrm{with} \, O_0 \equiv \mathbb{I}.
\end{equation}
The minimization can be rewritten in the case of a small (imaginary) time step $\Delta \tau$,
\begin{equation}
    \theta_\alpha' = \theta_\alpha - \Delta \tau \sum_\mu S^{-1}_{\alpha \mu} \frac{\partial E}{\partial \theta_\mu} \label{eq:update},
\end{equation}
where $E$ is the energy of the variational ansatz for a given hamiltonian $H$. The energy gradient can be expressed in terms of the logarithmic derivatives,
\begin{equation}
\frac{\partial E}{\partial \theta_\mu} = 2 \left({ \expval{O_\mu E_{loc}} - \expval{O_\mu}\expval{E_{loc}} }\right) = 2 \expval{\bar{O}_\mu \bar{E}_{loc}}.
\end{equation}
Here $E_{loc}(\bm{x})$ is defined as $\sum_{x'}\bra{\bm{x}} H \ket{\bm{x'}} \frac{\psi_{\theta}(\bm{x'})}{\psi_{\theta}(\bm{x})} $ -- measuring the ``local" energies associated with each spin configuration $\bm{x}$. This training paradigm is called stochastic reconfiguration (SR) \cite{Sorella} within the context of variational Monte Carlo. Typically this involves the inversion of an $(N_{params}, N_{params})$ matrix. This can be simplified to the inversion of an $(N_{samples}, N_{samples})$ matrix with the introduction of the neural tangent kernel (NTK). Using $T$ for the kernel,
\begin{equation}
T_{a b} = \sum_{\mu} \bar{O}_\mu(\bm{x_a}) \bar{O}_\mu(\bm{x_b}).
\end{equation}
Through a linear algebraic identity, Eq.~\eqref{eq:update} can be re-written,
\begin{equation}
    \theta_\alpha' = \theta_\alpha - 2 \Delta \tau \expval{\bar{O}_{\alpha} \expval{T_{ab} \bar{E}_{loc}}_a}_b.
\end{equation}
This application of the identity in this context is known as the kernel trick, and for a more thorough derivation handling complex paramters we direct the reader to Refs.~\cite{rende, minSR}. The NTK describes the $S$-matrix in function space, averaging over parameters rather than over samples. For infinite-width neural networks, analysis of NTK at initialization is sufficient to characterize training dynamics - a result that has been empirically shown for practical finite-sized networks as well \cite{NTK2}. Efficient estimation and inversion of the $S$ matrix is crucial for the standard approach of time-evolving variational states - time-dependent variational Monte Carlo (t-VMC) \cite{t-VMC1, t-VMC2}. Stochastic estimates of this matrix can suffer from bias when the ans\"atz represents a state with vanishing amplitudes in the computational basis. Such states can also cause Monte Carlo estimators to have low signal-to-noise ratio - defined as the ratio of the squared mean to the variance - thereby necessitating exponentially large number of samples at each training iteration. The spectrum of the estimated $S$ matrix is rank-deficient, and as a result the inversion in Eq.~\eqref{eq:update} becomes unreliable and requires regularization to ensure stability \cite{t-VMC3}.

Minimally entangled typical thermal states (METTS) are an ensemble representation of the finite-temperature equilibrium density matrix \cite{white}. A METTS simulation involves initializing and evolving a set of variational states ($\ket{\varphi_i}$) that, with appropriate importance weights, represent the thermal behavior of the system. Any set of variational states that are initialized to satisfy
\begin{equation}
    \sum_i^{N_{states}}\ket{\varphi_i}\bra{\varphi_i} \approx \mathbb{I} \label{eq:init},
\end{equation}
 and are equipped with imaginary-time evolution will work as an ensemble representation of the Gibbs state. However, the METTS approach involves initial states being classical product states with a Gaussian random weight for each local degree of freedom per site. This choice restricts the growth of the entanglement and is straightforward, especially for simulations with matrix product states \cite{Stoudenmire}. To calculate the thermal expectation values at an inverse temperature $\beta$, we imaginary-time evolve the initialized states to $\beta/2$,
\begin{equation}
    \ket{\varphi_i(\beta)} = e^{\frac{-\beta H}{2}} \ket{\varphi_i}.
\end{equation}
Interpreting $ \sum_i \ket{\varphi_i(\beta)}\bra{\varphi_i(\beta)} \approx e^{-\beta H}$, we can calculate thermal averages of operators from the expectation values in the individual states and the norm of each state,
\begin{equation}
    \expval{\mathcal{O}}_\beta \equiv \frac{\Tr[e^{-\beta H} \mathcal{O}]}{\Tr[e^{-\beta H}]} = \frac{\sum_i \braket{\varphi_i(\beta)}{\varphi_i(\beta)} \expval{\mathcal{O}}_i}{\sum_i \braket{\varphi_i(\beta)}{\varphi_i(\beta)}} \label{eq:avg}.
\end{equation}
The norm of each state serves as the importance weight for it, $Z_i(\beta)\equiv \braket{\varphi_i(\beta)}{\varphi_i(\beta)}$, and can be calculated efficiently by leveraging the imaginary time Schrodinger equation,
$\partial_\beta \ln Z_i(\beta) = -E_i (\beta)$,
\begin{equation}
    Z_i(\beta) = Z_i (0) \exp(- \int_0^\beta E_i(\beta ') d \beta') \label{eq:impWeights}.
\end{equation}
Finally, when using variational states parameterized by neural networks, the representational capacity of the network and the stochastic evaluation of gradients can induce errors in the imaginary time evolution. The authors of Ref.~\cite{hendry} suggest accounting for this by creating a mapping between the imaginary time $\tau$ and the inverse temperature $\beta$ for each state. $\tau$ is a directly determined by the learning rate $\Delta \tau$ and the number of training iterations. $\tau$ can be related to the inverse temperature through the mapping,
\begin{equation} 
    \beta_i(\tau) = - \int_0^{\tau} d\tau'  \frac{1}{\sigma_i^2} \frac{d E_i}{d \tau'}  \label{eq:beta}.
\end{equation}
\begin{figure}[t]
\centering
\includegraphics[width = 0.5\textwidth]{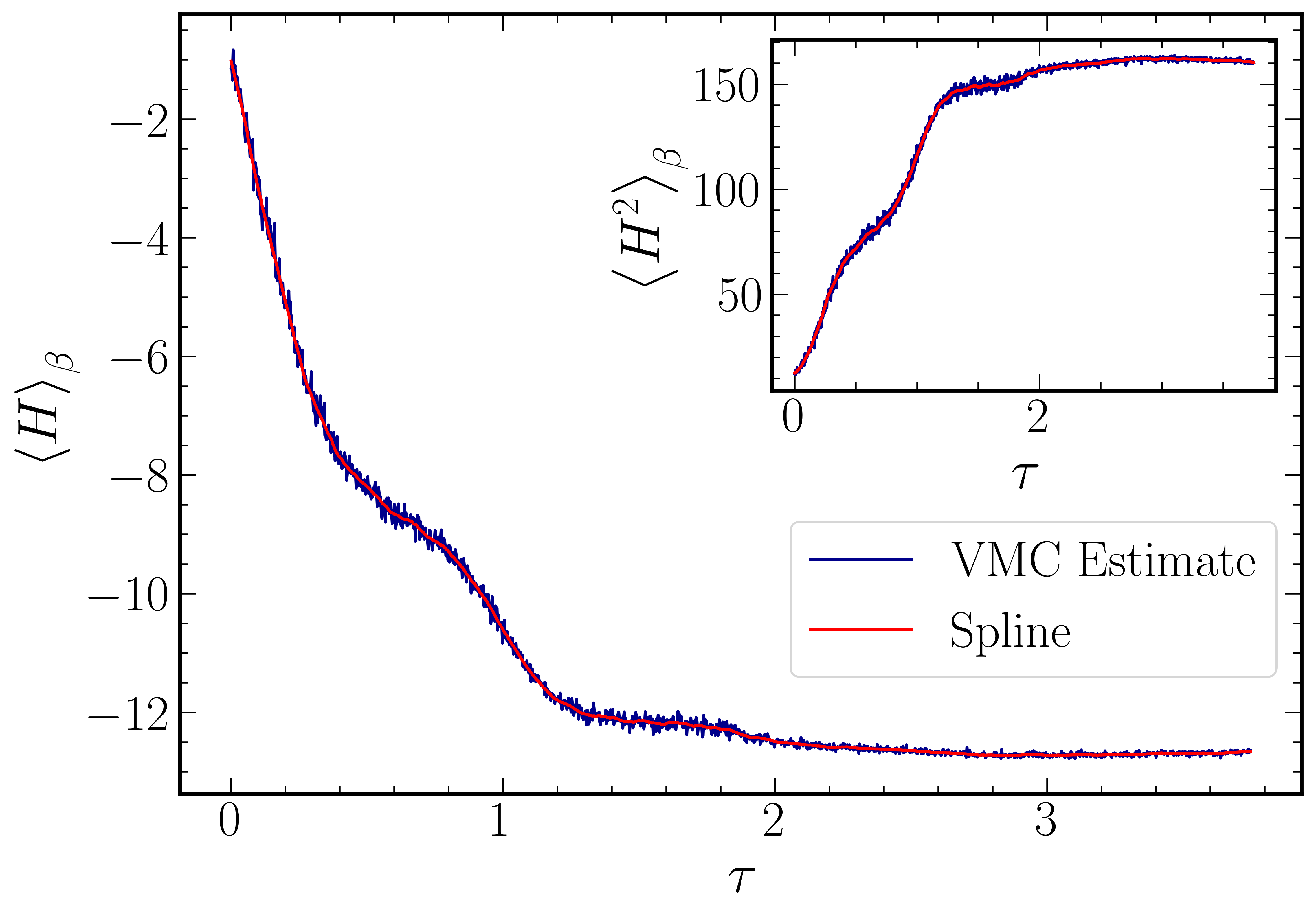} 
\centering
\caption{Numerical errors can compound while calculating the importance weights, $Z_i(\tau)$, and reparameterized inverse temperature, $\beta_i(\tau)$. This is mitigated by fitting VMC estimates of observables with splines.}
\label{fig:spl}
\end{figure}

This is motivated by the thermodynamic identity $dE/d\beta = -\sigma^2$, which is automatically enforced if $\beta_i$ is calculated as above. The reparametrization requires the calculation of the variance in the energy,
\begin{equation}
    \sigma_i^2 = \expval{H^2}_i - \expval{H}_i^2,
\end{equation} 
which comes at an additional cost proportional $N_{loc}^2$ where $N_{loc}$ is the number of terms in the Hamiltonian.

\section{Autoregressive implementation and results}\label{sec:results}
Autoregressive neural network can parameterize a variational state through conditional amplitudes via Eq.~\eqref{eq:psiJoint}.
In this work, we use a two-layer recurrent neural network (RNN) with an LSTM cell as the autoregressive ans\"atz \cite{lstm, rnnWavefns},
with the goal of simulating 
a spin-1/2 quantum XY chain with $N=20$ sites, 
\begin{equation}
    H = \frac{1}{2}\sum_{j = 1}^{N} \left[ \sigma_j^{x} \sigma_{j + 1}^{x} +   \sigma_j^{y} \sigma_{j + 1}^{y}  \right].
    \label{XYHam}
\end{equation}

The first layer of the RNN architechture (Fig. \ref{fig:ansatz}) has a hidden dimension of 10, the second layer has hidden dimension of 2, and we use fully complex parameters for all weights and biases. The output of the second layer is interpreted as logits: unnormalized conditional log amplitudes. This is motivated by two factors. One is that the calculation of the importance weights in Eq.~\eqref{eq:impWeights} implicitly assumes that the variational states are unnormalized. The second is the explicit normalization of autoregressive networks has been found to restrict their capability \cite{Booth}. Autoregressive sampling is still possible, as you can calculate the normalization constant across the degrees of freedom in the local Hilbert space. But the unnormalized amplitudes are used in the estimation of the gradients. An RNN ans\"atz can be explicitly initialized to classical product states by adding site-dependent bias terms. The recurrent cell's variables are set exactly to zeros. Reference \cite{hendry} suggests initializing them to Gaussian random variables, but we find that this affects performance, mainly at $\beta < 1$. 
\begin{figure}[!t]
\centering
\includegraphics[width = 0.5\textwidth]{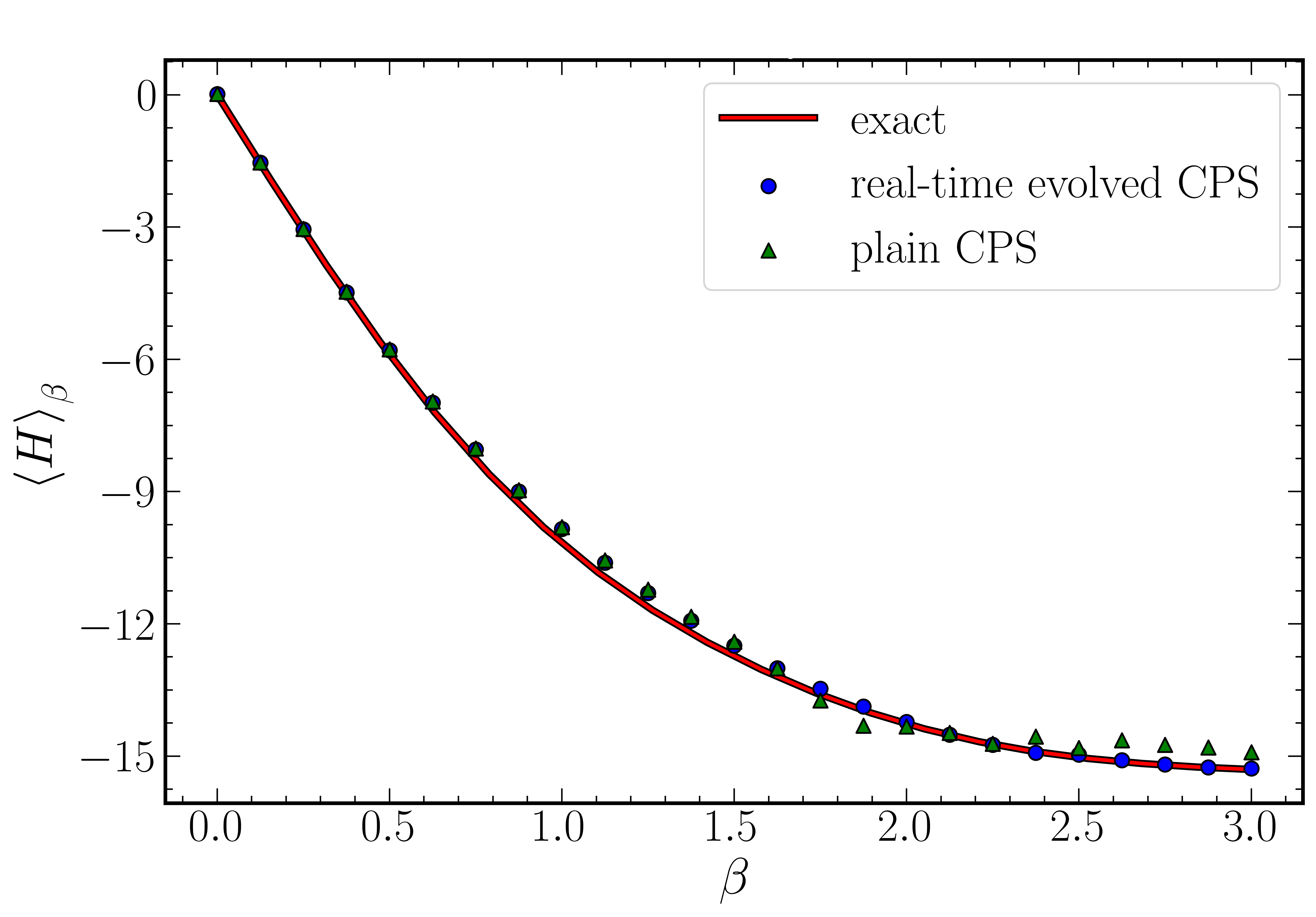}
\caption{Comparison of the thermal estimates of energy from autoregressive typical states initialized with and without the real time pre-evolution.}
\label{fig:main}
\end{figure}
We benchmark the performance of autoregressive typical thermal states on our XY model, Eq.~\eqref{XYHam}, which is exactly solvable through a mapping to free fermions \cite{exactSolution}.
An initial observation of our technique is that
the expectation value of the energy and its variance
accrue significant statistical variation (see Fig.~\ref{fig:spl}), which requires smoothing (e.g. by splines) as a function of $\tau$ in order to significantly reduce the proliferation of errors in the calculations outlines by Eqs. \eqref{eq:avg} and \eqref{eq:beta}. Particularly, in Eq.~\eqref{eq:beta} the integrand is poorly behaved due to fluctuations in variational estimates. As the RNN converges towards the ground state, the energy variance $\sigma_i^2$ tends to zero. When fluctuations lead to a vanishing denominator in $\sigma_i^2$, but a comparatively large $dE_i / d\tau$, the integrand blows up resulting in an inaccurate estimate of $\beta_i(\tau)$. We find that using a Savitzky-Golay filter on estimates of $\expval{H}$ and $\expval{H^2}$ makes the calculation in Eq.~\ref{eq:beta} more well-behaved. An essential detail is that these estimates are functions of $\tau$, but the thermal estimates from Eq.~\eqref{eq:avg} and Eq.~\eqref{eq:impWeights} require them as a function of $\beta$. This can be achieved by using a interpolation function that reparameterizes the functions of $\tau$ as functions of $\beta$, through the inversion of $\beta(\tau)$.

Another observation is that, 
RNN wave functions can exhibit unstable optimization, which is further enhanced in a stochastic reconfiguration scheme \cite{vanExpGradRnn, donatella, fidelity}. 
Hence, to perform typical thermal state simulations with RNN ans\"atze, we make two modifications to the procedure outlined in Section \ref{sec:background}. 
\begin{enumerate}
\item {\it The initial classical product states in METTS are replaced with classical product states evolved in real time.} 
\end{enumerate}
There are some physical motivations for selecting the initialized states as product states, originating from the low entanglement entropy of each imaginary-time evolved state \cite{white}. This is especially prominent when performing simulations with matrix product states. However, neural network wavefunctions can represent highly entangled states and the principle of estimating thermal observables with an ensemble of imaginary-time evolves states holds as long as Eq.~\eqref{eq:init} applies. We find that the RNNs initialized to product states struggle to reach low energies through imaginary time evolution - effectively getting stuck in local minima. Limitations with stochastic reconfiguration for neural networks that represent states close to product states have been studied before \cite{geoNQS, fidelity}. In such cases, the VMC estimate of the $S$ matrix from Eq.~\eqref{eq:S} becomes increasingly rank deficient and has a low signal-to-noise ratio, causing problems with the calculation of $S^{-1}$ in Eq.~\eqref{eq:update}. 
Viewed through the lens of the neural tangent kernel, poor conditioning of the this matrix at initialization will affect the performance of the RNNs throughout the simulation. Thus, we attempt to mitigate the problem by real-time evolving our initialized RNNs, prior to the imaginary-time evolution. The resultant improvement in convergence to the ground state is shown in Fig.~\ref{fig:main}. Since real-time evolution is unitary, the variational states still obey Eq.~\eqref{eq:init}. Interestingly, the real-time evolution we have used is also based on traditional TDVP using the S-matrix, and thus should be prone to the same inaccuracies. However, the goal is not to simulate exact time dynamics, but rather to evolve the initialized states with any unitary. So the errors induced from issues with SR in the real-time evolution step are not a roadblock.   

We continue with the second modification to the standard METTS procedure:

\begin{enumerate}
\setcounter{enumi}{1} 
\item {\it We stop imaginary time evolution of a variational state if}
\begin{equation*}
    \abs{\frac{d \beta_i}{d \tau}} > \Theta,
\end{equation*}
{\it where $\Theta$ serves as a threshold.  }
\end{enumerate}

This condition eliminates large oscillations in $\beta_i(\tau)$, which we observe for some of the initialized RNN states. VMC estimates are functions of $\tau$ and need to be reparameterized in terms of $\beta_i$ - requiring $\beta_i(\tau)$ to be monotonic. Therefore, these large oscillations can greatly bias thermal estimates calculated using Eq.~\eqref{eq:avg}. Intuitively, this condition can be interpreted as the elimination of energy spikes during training. Stable variational optimization should result in a smooth decrease of energy. However, RNNs can experience sudden jumps in the loss value (in our case, the energy) due to vanishing or exploding gradients \cite{eisenmann}. Another reason to use this threshold as a termination condition is that as the ans\"atz converges to the ground-state, the variance of the energy should tend to zero. This makes Eq.~\eqref{eq:beta} ill defined, even with the Savitzky-Golay smoothing, and thus it is a natural stopping point for the optimization. Without the thresholding, the thermal averages have huge deviations at all temperatures, caused by averaging over a few poorly evolved variational states.

\begin{figure*}[!t]
    \begin{subfigure}{0.47 \textwidth}
    \centering
    \subcaption{}
     \includegraphics[width = \textwidth]{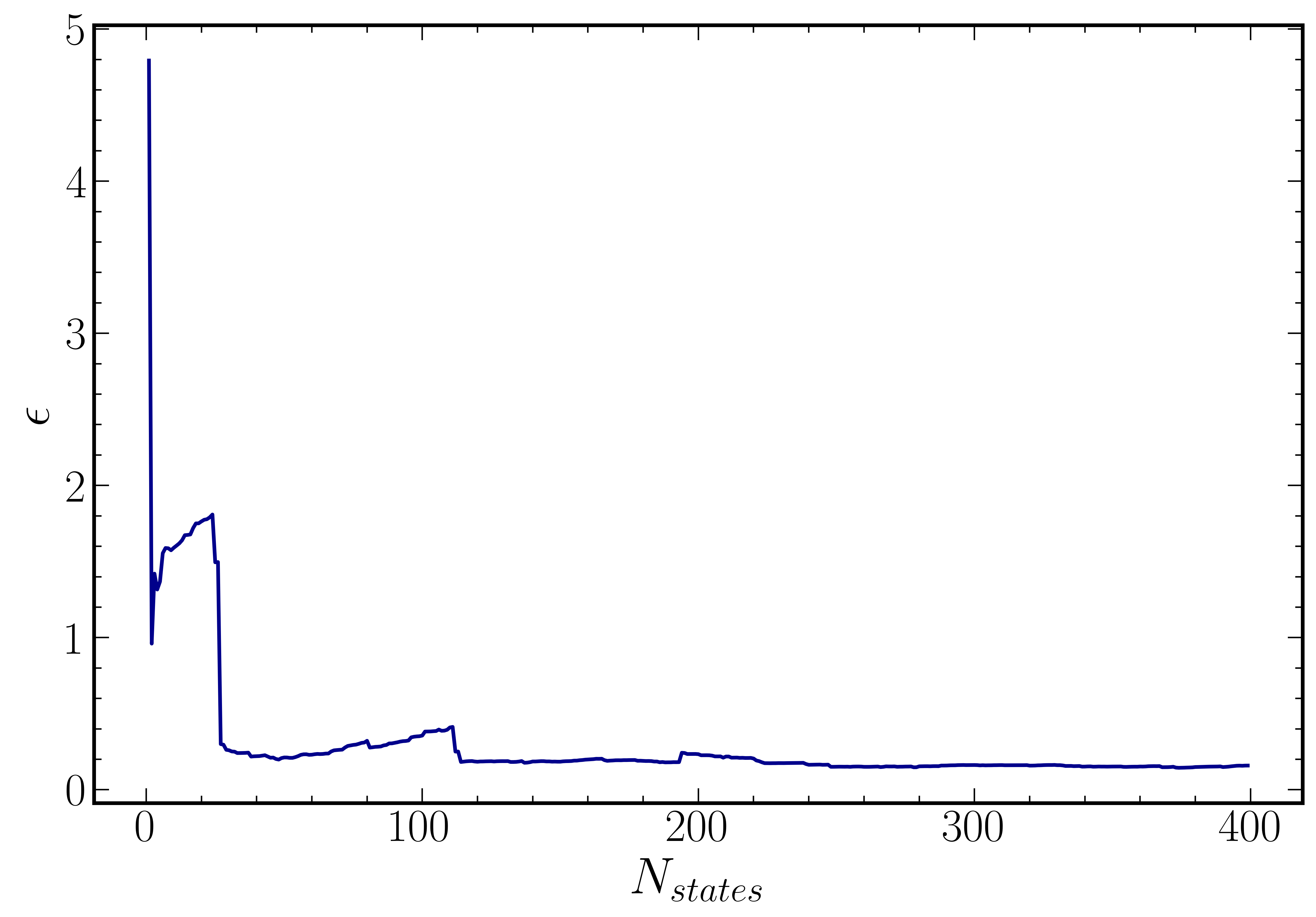}
    \end{subfigure}
    \hspace{7.5mm}
    \begin{subfigure}{0.47 \textwidth}
    \centering
    \subcaption{}
    \includegraphics[width = \textwidth]{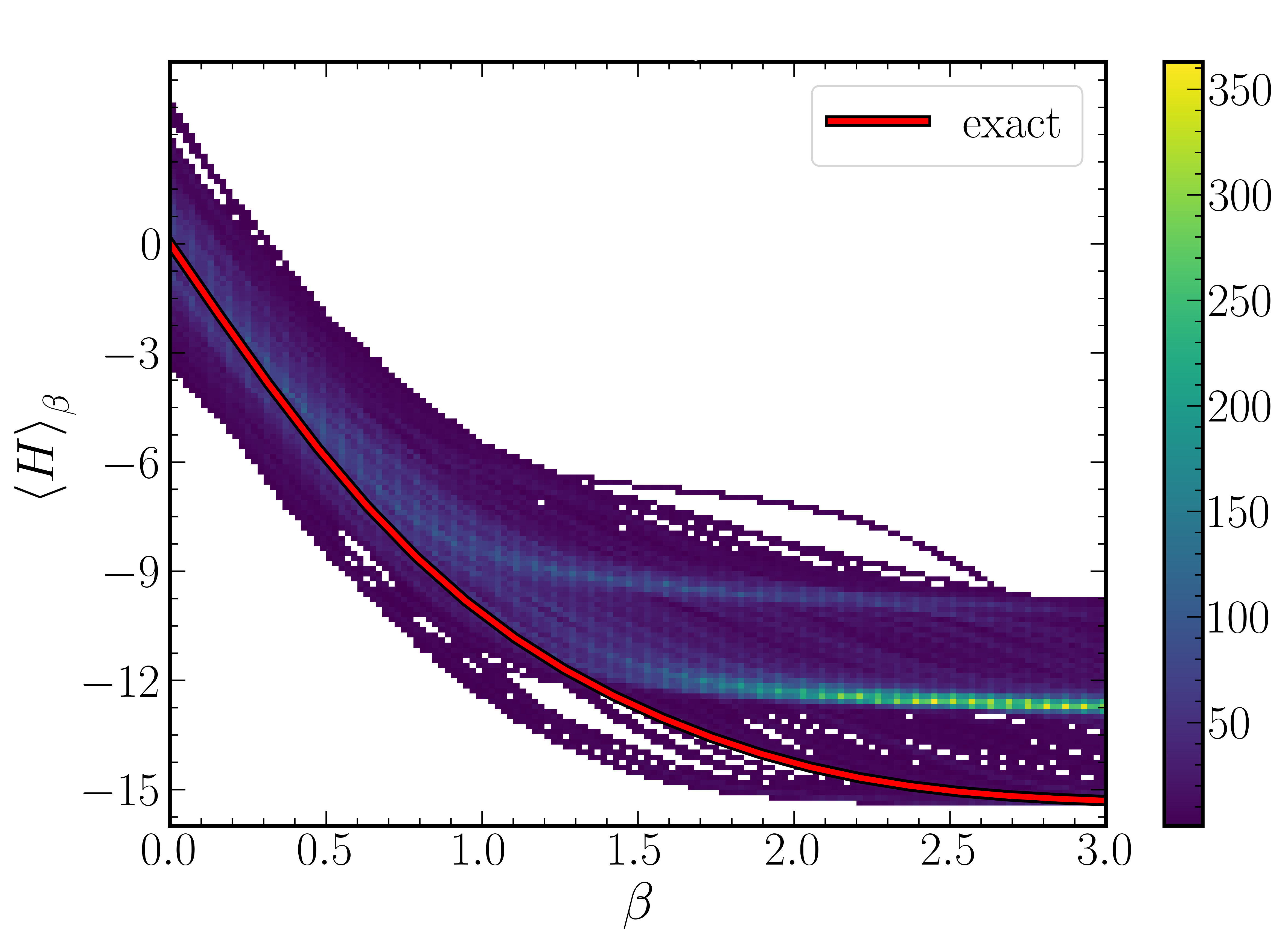}
    \end{subfigure}
    \caption{\textbf{(a)} Improvement with increasing number of variational states is nonuniform. \textbf{(b)} 2D heatmap of the energies of 400 variational states reparameterized as a function of $\beta$. The areas of high-density (represented through the colormap) show that most initialized states are stuck in local minima and do not reach the ground state.} 
    \label{fig:comparisons}
\end{figure*}

The performance of the simulation at high temperatures ($\beta \approx 0$) is controlled by the initialized states. In the traditional METTS algorithm, as the initialized states are simply CPS, the high temperature performance depends only on their statisitcs. Both the accurate thermal estimates (see Fig.~\ref{fig:main}) and the energy distribution at $\beta \ll 1$ (see Fig.~\ref{fig:comparisons}-b) indicate that the real-time pre-evolution does not bias the performance of the typical thermal state algorithm. Low temperature ($\beta \rightarrow\infty$) performance depends on the convergence of variational states to the ground state. This can be improved by increasing the expressiveness of the ansatz, using a lower learning rate (smaller $\Delta \tau$), imposing Hamiltonian symmetries, and many other techniques explored in ground-state studies with autoregressive networks. A crucial insight is that only a fraction of the initialized variational states need to approximately reach the ground state as $\beta \rightarrow \infty$. The reason for this is two-fold - the variational estimates of observables are reparameterized in terms of $\beta$ and the importance weights in \eqref{eq:impWeights} appropriately prioritize lower energy states. Thermal estimates at intermediate temperatures is most prone to inaccuracy (Fig.~\ref{fig:main}). To explore performance improvements from increasing the number of initialized states ($N_{states}$), we define the error in the energy, 
\begin{equation}
    \epsilon \equiv \int_1^{2.5} \abs{E_{estimate} - E_{exact}}~ d \beta .
\end{equation}
As Fig.~\ref{fig:comparisons}-a details, increasing the number of variational states does not provide a continuous improvement. Only a small fraction of the initialized states outperform the exact thermal energy \textemdash especially at intermediate $\beta$. When taking the average over variational states, the importance weights in Eq.~\eqref{eq:impWeights} prioritize states that reach lower energies. Thus, the discontinuous jumps can be explained as the inclusion of one of these ``good'' states. 
\section{Discussion}
In this work, we have presented a modification of the minimally entangled typical thermal states algorithm to accommodate autoregressive neural networks. These adjustments address challenges in simulating time dynamics with autoregressive models, and particularly recurrent neural network wavefunctions, which are sensitive to initialization and prone to numerical instabilities during gradient-based training. By entangling the product states with unitary evolution, our approach improves initialization and enables accurate estimation of thermal observables. Additionally, imposing a cutoff to remove divergences during optimization leverages the ensemble nature of METTS—while optimization spikes may affect individual trajectories, the overall finite-temperature behavior of the ensemble remains intact.

As improvement from increasing the number of initialized states seems to saturate, an interesting parallel avenue for improvement is better initialization of the variational states. It is generally accepted in the machine learning community that sophisticated initialization techniques can have an outsized effect on the performance of neural networks. We have explored one method to improve intialization from classical product states by using real-time evolution. Clearly, the exploration of how different unitaries or evolution times affect performance is an area for further study. Studying ensembles of RNNs under different hyperparamters, such as hidden layer size, differring non-linearities and initialization schemes, etc.~could provide insight into the performance of this typical thermal state algorithm.

In conclusion, there is still much work to do to develop generative model strategies for simulating finite-temperature quantum states. 
In particular, it would be interesting to devise a more direct comparison of the performance of autoregressive typical thermal states to METTS implemented with energy-based models.
Such energy-based ansätze rely on Markov-chain Monte Carlo techniques for optimization, which can results in samples with long autocorrelation times
in many situations familiar to physicists.
Whether the capability of autoregressive models to perform exact sampling provides any advantage in these situations is still an open question.  Further study of cases with long autocorrelation times, such as at finite-temperature phase transitions, would be interesting.
Autoregressive models may also provide value in cases where having a normalized distribution $p({\bm x})$ is required, such as in the calculation of entropies or free energies.
Looking forward, the hope is that with continued progress such as this, autoregressive quantum states can replicate even a fraction of the success at scaling as observed by their industry counterparts, such as those found in today's large language models.

\section{Acknowledgements}
We thank Juan Carrasquilla, Schuyler Moss, Yi-Hong Teoh, Roeland Wiersema, Rimika Jaiswal, and Gurpahul Singh for helpful discussions. The simulations in this work were carried out using NetKet \cite{netket2:2019, netket3:2022}, which relies on Jax \cite{jax2018github} and MPI4Jax \cite{mpi4jax:2021}. Computational resources were provided by the Shared Hierarchical Academic Research Computing Network (SHARCNET) and
the Digital Research Alliance of Canada.
We acknowledge support from the
Natural Sciences and Engineering Research Council of Canada (NSERC) and the
Perimeter Institute. L.B. was supported by the NSF CMMT program under Grants No. DMR-2419871, and the Simons Collaboration on Ultra-Quantum Matter, which is a grant from the Simons Foundation (Grant No. 651440).  Research at Perimeter Institute is supported in part by the Government of Canada through
the Department of Innovation, Science and Economic
Development Canada and by the Province of Ontario
through the Ministry of Economic Development, Job
Creation and Trade.  

\bibliography{references}

\end{document}